# Development of a digital tool for monitoring the behaviour of pre-weaned calves using accelerometer neck-collars.

Oshana Dissanayake[1,3], Sarah E. McPherson[3,4,5], Joseph Allyndrée[2,3], Emer Kennedy[3,4], Pádraig Cunningham[1], Lucile Riaboff[1,3,6]

[1] School of Computer Science, University College Dublin, Ireland
[2] School of Maths and Stats, University College Dublin, Ireland
[3] VistaMilk SFI Research Centre
[4] Teagasc, Animal & Grassland Research and Innovation Centre, Cork, Ireland
[5] Animal Production Systems Group, Wageningen University & Research, Wageningen, The Netherlands
[6] GenPhySE, Université de Toulouse, INRAE, ENVT, 31326, Castanet-Tolosan, France
oshana.dissanayake@ucdconnect.ie

## Abstract

Automatic monitoring of calf behaviour is a promising way of assessing animal welfare from their first week on farms. This study aims to (i) develop machine learning models from accelerometer data to classify the main behaviours of pre-weaned calves and (ii) set up a digital tool for monitoring the behaviour of pre-weaned calves from the models' prediction. Thirty pre-weaned calves were equipped with a 3-D accelerometer attached to a neck-collar for two months and filmed simultaneously. The behaviours were annotated, resulting in 27.4 hours of observation aligned with the accelerometer data. The time-series were then split into 3 seconds windows. Two machine learning models were tuned using data from 80% of the calves: (i) a Random Forest model to classify between active and inactive behaviours using a set of 11 hand-craft features [model 1] and (ii) a RidgeClassifierCV model to classify between lying, running, drinking milk and other behaviours using ROCKET features [model 2]. The performance of the models was tested using data from the remaining 20% of the calves. Model 1 achieved a balanced accuracy of 0.92. Model 2 achieved a balanced accuracy of 0.84. Behavioural metrics such as daily activity ratio and episodes of running, lying, drinking milk, and other behaviours expressed over time were deduced from the predictions. All the development was finally embedded into a Python dashboard so that the individual calf metrics could be displayed directly from the raw accelerometer files.

## Introduction

Adopting appropriate animal husbandry practices offers many benefits; it enhances livestock welfare, increases farm productivity, and improves the quality of the resulting agricultural products (Blokhuis et al., 2003). Stress in animals can lead to weight loss, exhaustion, disease, and even death, particularly in young animals that are more vulnerable. Improving calf welfare is thus crucial to prevent physiological changes and death, especially in prolonged stress conditions (Koknaroglu and Akunal, 2013). Furthermore, prioritizing welfare will help society to trust farm practices more, as the welfare of the animals is a primary concern of the customers (Cardoso et al., 2016). Calf behaviour changes, such as altered feeding patterns, reduced playing time, social isolation, and decreased movement, can indicate health concerns or environmental stressors (Mahendran et al., 2023; Nikkhah and Alimirzaei, 2023). Monitoring these behaviours to detect any deviation is thus a key way to detect disturbances in young calves (stressful events, diseases, etc.) (Dissanayake et al., 2022). However, continuous behaviour monitoring is required, which is impractical with human observation. Livestock behaviour can be obtained from onboard accelerometer sensors in different species and systems to get various behaviours. Accelerometers are adaptable and versatile and can function for an extended period, making them a popular choice in livestock behaviour monitoring (Riaboff et al., 2022).

In the literature, accelerometer signal segmentation, hand-crafted feature calculation, and machine learning modelling are commonly used for animal behavior classification. Studies reported high classification performance in dairy cattle, particularly with Random Forest (RF), eXtreme Gradient Boosting and Support Vector Machine (Riaboff et al., 2022). However, only one study has attempted to do so on dairy calves (Carslake et al., 2020). Furthermore, the models' ability to classify a wide range of behaviours with good genericity from one animal to another is still challenging (Riaboff et al., 2022). This calls for applying new techniques that have proved their worth in other time series classification problems. For example, Random Convolutional Kernel Transform (ROCKET; (Dempster et al., 2020)) combined with RidgeClassifierCV (RCV) can capture diverse temporal patterns without domain-specific knowledge to distinguish patterns and underlying structures within the data. They thus appear as a promising alternative to the hand-crafted features typically used in the field, particularly when the time series show patterns specific to their class label.

Therefore, the primary goal of this study is to (i) classify calf behaviour as active and inactive, incorporating hand-crafted features, (ii) classify a set of four calf behaviours incorporating ROCKET features, and (iii) develop a behaviour metric dashboard to utilising behaviour durations, behaviour amounts based on the time of the day etc., incorporating these active/inactive and calf behaviour classification models.

**Materials and methods**

<u>Active/inactive states and calf behaviour classification</u>

The complete methodology followed is shown in Figure 1.

The trial for the data collection was carried out at Teagasc Moorepark Research Farm (Fermoy, Co. Cork, Ireland; 50◦07′N; 8◦16′W) using 30 Holstein Friesian and Jersey pre-weaned calves housed in grouped pens from 21/01/2022 to 05/04/2022 (experimental procedures in accordance with the European Union (Protection of Animals Used for Scientific Purposes) Regulations 2012 (S.I. No. 543 of 2012); ethical approval: TAEC2021–319). Calves were equipped with a tri-axes accelerometer sensor attached to a neck collar (Axivity LTD[1]; sampling rate: 25 Hz) and filmed simultaneously. Calves' behaviour was then annotated from the videos using

---

[1] https://axivity.com/product/ax3

Behavioural Observation Research Interactive Software (BORIS; Friard and Gamba, 2016) so that each calf was observed for at least 15 minutes. Detailed explanation on the trial, amount of data collected and annotated behaviours can be found at: {section 2.1: *http://arxiv.org/abs/2404.18159*}. Next, the raw accelerometer time-series (X, Y, Z) and additional time-series (magnitude, ODBA, VeDBA, pitch, roll) were segmented (3-seconds windows; 50% overlap) and aligned with the corresponding observed behaviour (gold standard). 88 hand-crafted features and ROCKET features were then calculated for each of the segmented windows. The method applied for time-series calculation, segmentation and features can be found at: {section 2.2: *http://arxiv.org/abs/2404.18159*}. The dataset was then split into a ratio of 20:80, thus resulting in 6 calves out of 30 to be included in the test set and 24 calves in the training set. Model tuning was performed with 10-fold repeated hold-out validation using a grid search process. Balanced accuracy, sensitivity, specificity and precision were then calculated using the test set to evaluate model performance. Each fold was composed of 5 validation calves and 19 train calves (20:80 split out of 24 train calves). RF classifier was used to classify active (the calf is walking, playing, drinking, eating, grooming itself, interacting with other calves or its environment) *versus* inactive behaviours (calf is standing still or lying) from the hand-crafted features. RCV classifier was used to fit a model to classify lying, running, drinking milk and a single class called "other" merging the rest of the annotated behaviours. The method applied for splitting the dataset, model tuning, model testing and evaluation is described in detail in {sections 2.3, 2.4 and 2.5: *http://arxiv.org/abs/2404.18159*}.

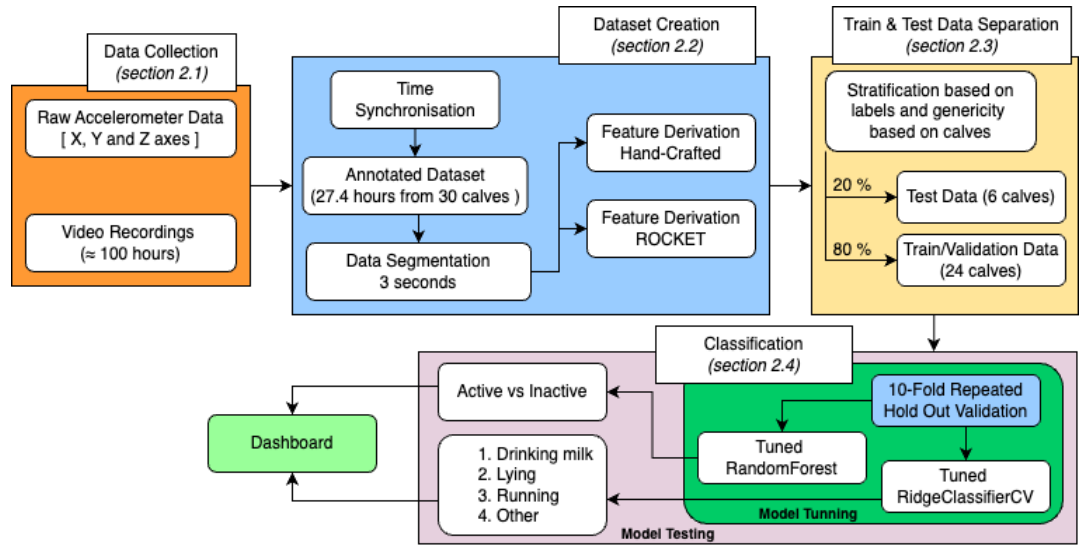

**Figure 1**: Pipeline

Dashboard creation

A Python-based (Dash[2]: Python framework for building web applications, utilizing popular libraries like Pandas, Plotly, and Flask) dashboard was developed to calculate key behaviour metrics using RF and RCV model predictions from 2 months of accelerometer data collected from 47 calves.

A user can upload a raw accelerometer file (.cwa) to the application, and it automatically converts the file into a readable format. This data is then segmented into 3-second windows. Hand-crafted and ROCKET features are calculated for each window. The pre-trained RF and RCV classification models then classify these windows into active/inactive and lying, running, drinking milk, and other. Finally, the predictions are used to derive useful metrics on calf behaviour (see Table 1).

**Table 1**: Information displayed in the dashboard.

| Information | Metrics | Definition |
|---|---|---|
| Activity over time | Active/Inactive ratio over time | Ratio between the active and inactive behaviour proportions per hour. |
| | Running proportion over time | Ratio of running behaviour against other considered behaviours per hour. |
| | Lying proportion over time | Ratio of lying behaviour against other considered behaviours per hour. |
| | Drinking milk proportion over time | Ratio of drinking milk behaviour against other considered behaviours per hour. |
| | Other behaviours proportion over time | Ratio of other behaviours against the rest of the considered behaviours per hour. |
| Activity summary | Total active vs inactive ratio | The proportion of active and inactive behaviours within a given period of time. |
| | Total behaviour ratio | The proportion of each behaviour within a given period of time. |
| | Day Night active ratio | The proportion of active behaviour during the daytime (6 am to 8 pm ) and night-time (8 pm - 6 am) within a given period of time. |
| | Day Night inactive ratio | The proportion of inactive behaviour during the daytime (6 am to 8 pm ) and night-time (8 pm - 6 am) within a given period of time. |

3. **Results and Discussion**

Active / Inactive classification

RF with hand-crafted features performed exceptionally well in classifying the data as active and inactive (Figure 2), with a balanced accuracy of 0.92, on calves not seen during model training. The inactive class has achieved the highest precision (0.96), and the active class has the highest recall (0.94).

---
[2] https://dash.plotly.com/

Behaviour classification

RCV with ROCKET features has performed well with a balanced accuracy of 0.84. Running behaviour has the best recall of 0.99, while the lying behaviour has the best precision of 0.89. The lowest recall was recorded for the other behaviour class (0.73), and the lowest precision was for the drinking milk behaviour (0.55). Drinking milk behaviour has been mostly confused with the other behaviour class (see Figure 3). Other behaviour class houses 21 different behaviours, thus making it more complex and more prone to be confused with the main behaviours. The "other" behaviour class has been mostly misclassified as the drinking milk or lying.

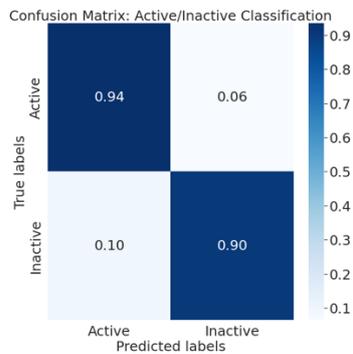

**Figure 2**: Confusion matrix for active and inactive classification.

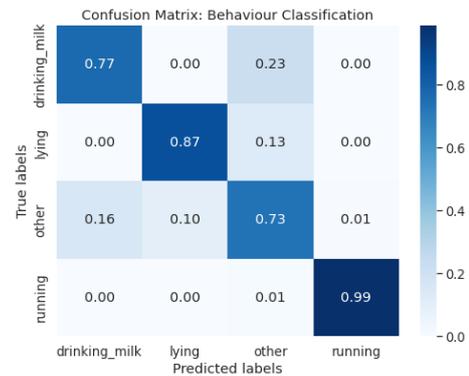

**Figure 3**: Confusion matrix for behaviour classification.

Dashboard

The dashboard provides the ability to filter the data by calf ID and date and displays information about the calf, such as breed, birth date, coat colour, and housed pen number (Figure 4). For each selected time period and for each calf, a set of outputs has been proposed to display the behavioural metrics (Table 1):

- **Graphs**: Plots of the hourly activity ratio have been proposed to follow the variations within the day (Figure 4); pie charts have been used to show the total activity *versus* inactive ratio and the ratio between the 4 behaviours during the selected period (Figure 5).

- **Tables**: Tables were created so that the active ratio per hour, as well as the running, lying, drinking milk, and other behaviour proportions per hour, could be extracted for further analysis (Figure 6). The prediction for each accelerometer timestamp (25 Hz) for the RF and RCV models is also available in a table format.

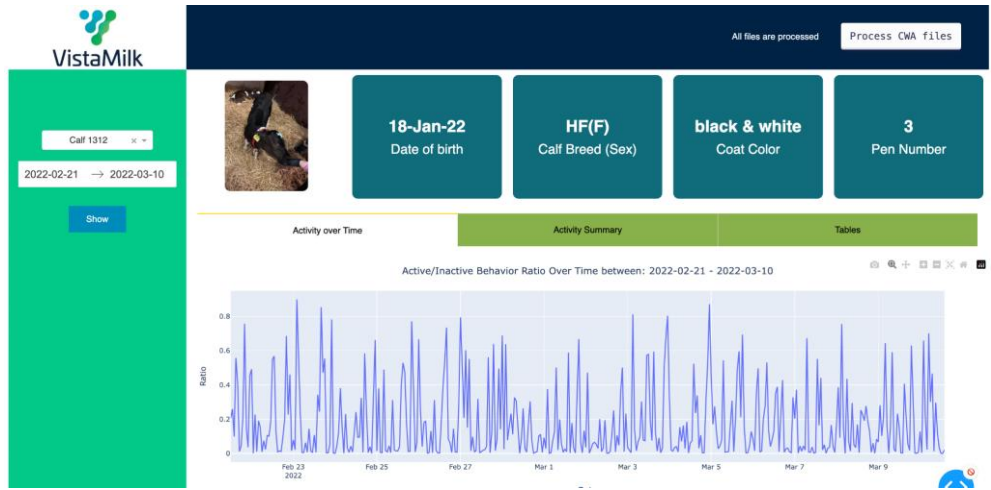

**Figure 4**: The landing page of the dashboard. Displays the functions to convert CWA to CSV, data filter options, calf information and Tab 01

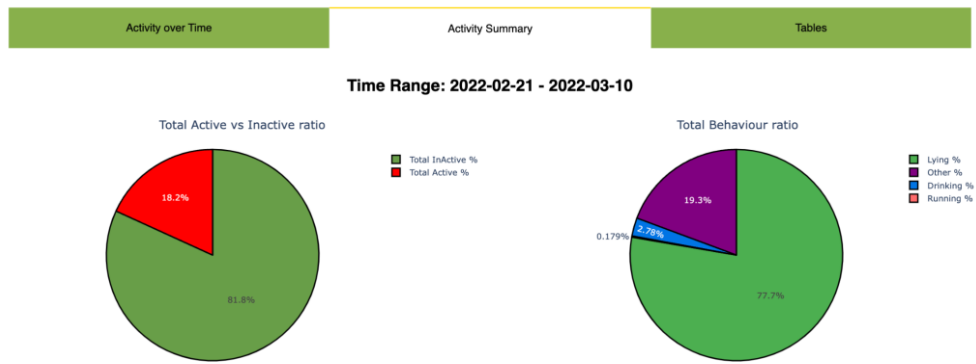

**Figure 5**: Tab 02 – Activity Summary

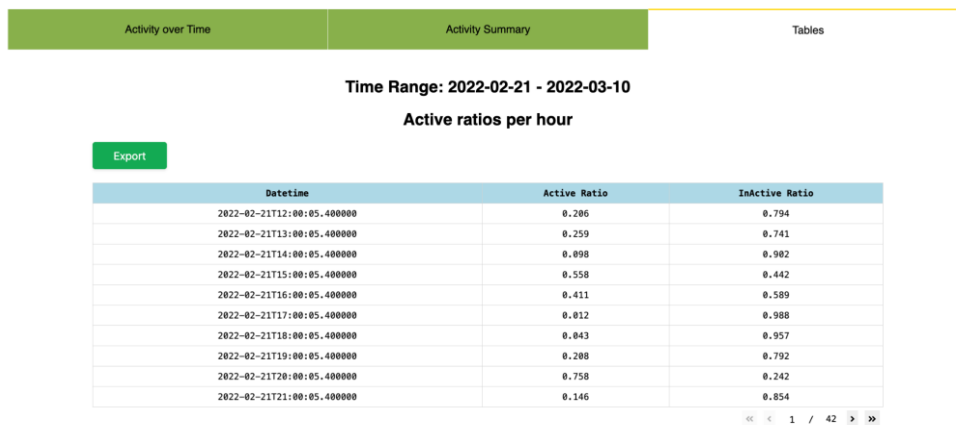

**Figure 06**: Tab 03 - Tables

Discussion

Improved calf welfare benefits farms by reducing disease prevalence, lowering treatment costs, reducing deaths, and improving product value. However, few studies have explored automatic monitoring of calf welfare, with accelerometer sensors showing promise in detecting deviations linked to stressful events or diseases.

The study aimed to create two classification models from accelerometer data: RF model with hand-crafted features to distinguish between active and inactive states and RCV model with ROCKET features to distinguish between drinking milk, running, grooming and the rest of the behaviours. Hand-crafted features effectively capture movement data distribution, while ROCKET features efficiently capture dynamic patterns. RF models yield better accuracy due to their ability to prevent overfitting. RCV can handle high dimensional data and use cross-validation to optimise regularisation strength, ensuring the model is neither too complex nor too simple, enhancing its generalisation capabilities to new data.

The performance of drinking milk is lower than running and lying behaviours due to its less variable nature and fewer discriminative features. The "other" class, which combines 21 different behaviours, does not show up well compared to running and lying behaviours due to its varied nature. This makes it difficult for the model to distinguish this class and create unique and robust identification. Overall, the classification model performs well, with acceptable class accuracies for drinking milk and other class and high accuracies for lying and running classes. This extensive range of behaviours provides valuable insights into sensitive areas, making it one of the first comprehensive studies to automatically monitor pre-weaned calves' behaviour from accelerometer data.

The study's primary goal was to create a user-friendly interface for collecting key calf behaviour metrics from the model's predictions, making the accelerometer data interpretable by animal scientists or farmers. The dashboard offers continuous and individual monitoring of calf behaviour metrics over a given period, including active/inactive behaviour ratio, running, lying, drinking milk, other behaviour proportions over time, total active versus inactive ratio, day/night active ratio, etc. The dashboard uses plots, pie charts, and tables to interpret the data efficiently. This allows monitoring behaviour metrics in different scenarios, such as weaning, dehorning, and relocation, and identifying individual deviations. An automatic, real-time collection of accelerometer data would encourage quick response to sick animals, improve decision-making, monitor compliance with established protocols, and develop effective strategies.

**Conclusion**

The study aimed to develop a behaviour metric dashboard using two behaviour classification models from accelerometer data. Hand-crafted accelerometer features with a Random Forest classifier were used to classify behaviour into active and inactive. ROCKET features with a RidgeClassifierCV were used to classify behaviours into drinking milk, running, lying and other. Both models performed with a balanced accuracy of 0.92 and 0.84, respectively. Overall, all the classes were accurately predicted, even if the model performed less well for drinking milk and other classes. The dashboard aims to make it a user-friendly interface for collecting key metrics of calf behaviour from the model's predictions, making the accelerometer data easily interpretable. A Python-based dashboard based on these models provides insight into calf behaviour, including active/inactive proportions over unit time and behaviour proportions over day and night.

**Acknowledgements**

This publication has emanated from research conducted with the financial support of SFI and the Department of Agriculture, Food and Marine on behalf of the Government of Ireland to the VistaMilk SFI Research Centre under Grant Number 16/RC/3835.